\documentstyle[10pt,epsfig,dp_delphititle,float,hangcaption,xspace,amssymb,
amsfonts,amsmath,amsthm,cite,graphicx,coll]{dp_delphi}
\makeindex
\def\DpPaperGroup{PH-EP}
\def\DpPaperRef{2006-023}
\def\DpDate{20 June 2006}
\def\DpAuthors{DELPHI Collaboration}
\def\DpSubmit{(Accepted by Eur. Phys. J. C)}
\def\DpTitle{{Search for a fourth generation  $\boldsymbol{\rm b'}$-quark at LEP-II at 
$\boldsymbol{\sqrt{s}=196-209}$~GeV}}
\def\DpComment{}
\def\DpEMail{}

\newcommand{\bbllnn}{${\mathrm b\bar b l^+ l^- \nu \bar\nu}$\xspace}
\newcommand{\bdbbllnn}{$\boldsymbol{\mathrm b\bar b l^+ l^- \nu \bar\nu}$\xspace}
\newcommand{\bbqqnn}{${\mathrm b\bar b q\bar q \nu \bar\nu}$\xspace}
\newcommand{\bdbbqqnn}{$\boldsymbol{\mathrm b\bar b q\bar q \nu \bar\nu}$\xspace}
\newcommand{\bbqqqq}{${\mathrm b\bar b q\bar q q\bar q}$\xspace}
\newcommand{\bdbbqqqq}{$\boldsymbol{\mathrm b\bar b q\bar q q\bar q}$\xspace}
\newcommand{\ccqqln}{${\mathrm c\bar c q\bar q l^+\nu}$\xspace}
\newcommand{\bdccqqln}{$\boldsymbol{\mathrm c\bar c q\bar q l^+\nu}$\xspace}
\newcommand{\ccqqqq}{${\mathrm c\bar c q\bar q q\bar q}$\xspace}
\newcommand{\bdccqqqq}{$\boldsymbol{\mathrm c\bar c q\bar q q\bar q}$\xspace}
\newcommand{\bbz}{${\mathrm b' \to bZ}$\xspace}
\newcommand{\brbz}{$BR({\mathrm b' \to bZ})$\xspace}
\newcommand{\bdbrbz}{$\boldsymbol{BR({\mathrm b' \to bZ})}$\xspace}
\newcommand{\bcw}{${\mathrm b' \to cW}$\xspace}
\newcommand{\brcw}{$BR({\mathrm b' \to cW})$\xspace}
\newcommand{\bdbrcw}{$\boldsymbol{BR({\mathrm b' \to cW})}$\xspace}
\newcommand{\bl}{${\mathrm b'}$\xspace}
\newcommand{\blq}{${\mathrm b'}$-quark\xspace}
\newcommand{\mblcem}{$m_{\rm b'}=100$~GeV$/c^2$\xspace}

\begin{document}

\makeatletter
\input{coll.sty}
\makeatother

\begin{titlepage}
\pagenumbering{roman}

\CERNpreprint{\DpPaperGroup}{\DpPaperRef}   
\date{{\small\DpDate}}                      
\title{\DpTitle}                            
\address{\DpAuthors}                        

\begin{shortabs}                            
\noindent
A search for the pair production of fourth generation
$b'$-quarks was performed using data taken by the DELPHI detector
at LEP-II. The analysed data were collected at centre-of-mass energies
ranging from 196 to 209~GeV, corresponding to an integrated luminosity
of 420~pb$^{-1}$. No evidence for a signal was found.
Upper limits on \brbz
and \brcw were
obtained for ${\mathrm b'}$ masses ranging from 96 to
103~GeV$/c^2$.
These limits, together with the theoretical branching ratios predicted
by a sequential four generations model, were used to constrain
the value of $R_{CKM}=|\frac{V_{\mathrm cb'}}{V_{\mathrm tb'}V_{\mathrm tb}}|$, where
$V_{\mathrm cb'}$, $V_{\mathrm tb'}$ and $V_{\mathrm tb}$ are elements of the extended 
CKM
matrix.

\end{shortabs}

\vfill

\begin{center}
\DpSubmit \ \\          
\DpComment \ \\
\DpEMail \ \\
\end{center}

\vfill
\clearpage

\headsep 10.0pt

\addtolength{\textheight}{10mm}
\addtolength{\footskip}{-5mm}
\begingroup
%
\newcommand{\DpName}[2]{\hbox{#1$^{\ref{#2}}$},\hfill}
\newcommand{\DpNameTwo}[3]{\hbox{#1$^{\ref{#2},\ref{#3}}$},\hfill}
\newcommand{\DpNameThree}[4]{\hbox{#1$^{\ref{#2},\ref{#3},\ref{#4}}$},\hfill}
\newskip\Bigfill \Bigfill = 0pt plus 1000fill
\newcommand{\DpNameLast}[2]{\hbox{#1$^{\ref{#2}}$}\hspace{\Bigfill}}

%
\footnotesize
\noindent
\DpName{J.Abdallah}{LPNHE}
\DpName{P.Abreu}{LIP}
\DpName{W.Adam}{VIENNA}
\DpName{P.Adzic}{DEMOKRITOS}
\DpName{T.Albrecht}{KARLSRUHE}
\DpName{R.Alemany-Fernandez}{CERN}
\DpName{T.Allmendinger}{KARLSRUHE}
\DpName{P.P.Allport}{LIVERPOOL}
\DpName{U.Amaldi}{MILANO2}
\DpName{N.Amapane}{TORINO}
% ...
\normalsize
\endgroup
\newpage
\titlefoot{Department of Physics and Astronomy, Iowa State
     University, Ames IA 50011-3160, USA
    \label{AMES}}
\titlefoot{IIHE, ULB-VUB,
     Pleinlaan 2, B-1050 Brussels, Belgium
    \label{BRUSSELS}}
\titlefoot{Physics Laboratory, University of Athens, Solonos Str.
     104, GR-10680 Athens, Greece
    \label{ATHENS}}
\titlefoot{Department of Physics, University of Bergen,
     All\'egaten 55, NO-5007 Bergen, Norway
    \label{BERGEN}}
\titlefoot{Dipartimento di Fisica, Universit\`a di Bologna and INFN,
     Via Irnerio 46, IT-40126 Bologna, Italy
    \label{BOLOGNA}}
\titlefoot{Centro Brasileiro de Pesquisas F\'{\i}sicas, rua Xavier Sigaud 150,
     BR-22290 Rio de Janeiro, Brazil
    \label{BRASIL-CBPF}}
\titlefoot{Inst. de F\'{\i}sica, Univ. Estadual do Rio de Janeiro,
     rua S\~{a}o Francisco Xavier 524, Rio de Janeiro, Brazil
    \label{BRASIL-IFUERJ}}
\titlefoot{Coll\`ege de France, Lab. de Physique Corpusculaire, IN2P3-CNRS,
     FR-75231 Paris Cedex 05, France
    \label{CDF}}
\titlefoot{CERN, CH-1211 Geneva 23, Switzerland
    \label{CERN}}
\titlefoot{Institut de Recherches Subatomiques, IN2P3 - CNRS/ULP - BP20,
     FR-67037 Strasbourg Cedex, France
    \label{CRN}}
\titlefoot{Now at DESY-Zeuthen, Platanenallee 6, D-15735 Zeuthen, Germany
    \label{DESY}}
\titlefoot{Institute of Nuclear Physics, N.C.S.R. Demokritos,
     P.O. Box 60228, GR-15310 Athens, Greece
    \label{DEMOKRITOS}}
\titlefoot{FZU, Inst. of Phys. of the C.A.S. High Energy Physics Division,
     Na Slovance 2, CZ-180 40, Praha 8, Czech Republic
    \label{FZU}}
\titlefoot{Dipartimento di Fisica, Universit\`a di Genova and INFN,
     Via Dodecaneso 33, IT-16146 Genova, Italy
    \label{GENOVA}}
\titlefoot{Institut des Sciences Nucl\'eaires, IN2P3-CNRS, Universit\'e
     de Grenoble 1, FR-38026 Grenoble Cedex, France
    \label{GRENOBLE}}
\titlefoot{Helsinki Institute of Physics and Department of Physical Sciences,
     P.O. Box 64, FIN-00014 University of Helsinki, 
     \indent~~Finland
    \label{HELSINKI}}
\titlefoot{Joint Institute for Nuclear Research, Dubna, Head Post
     Office, P.O. Box 79, RU-101 000 Moscow, Russian Federation
    \label{JINR}}
\titlefoot{Institut f\"ur Experimentelle Kernphysik,
     Universit\"at Karlsruhe, Postfach 6980, DE-76128 Karlsruhe,
     Germany
    \label{KARLSRUHE}}
\titlefoot{Institute of Nuclear Physics PAN,Ul. Radzikowskiego 152,
     PL-31142 Krakow, Poland
    \label{KRAKOW1}}
\titlefoot{Faculty of Physics and Nuclear Techniques, University of Mining
     and Metallurgy, PL-30055 Krakow, Poland
    \label{KRAKOW2}}
\titlefoot{Universit\'e de Paris-Sud, Lab. de l'Acc\'el\'erateur
     Lin\'eaire, IN2P3-CNRS, B\^{a}t. 200, FR-91405 Orsay Cedex, France
    \label{LAL}}
\titlefoot{School of Physics and Chemistry, University of Lancaster,
     Lancaster LA1 4YB, UK
    \label{LANCASTER}}
\titlefoot{LIP, FCUL, IST, CFCUC - Av. Elias Garcia, 14-$1^{o}$,
     PT-1000 Lisboa Codex, Portugal
    \label{LIP}}
\titlefoot{Department of Physics, University of Liverpool, P.O.
     Box 147, Liverpool L69 3BX, UK
    \label{LIVERPOOL}}
\titlefoot{Dept. of Physics and Astronomy, Kelvin Building,
     University of Glasgow, Glasgow G12 8QQ
    \label{GLASGOW}}
\titlefoot{LPNHE, IN2P3-CNRS, Univ.~Paris VI et VII, Tour 33 (RdC),
     4 place Jussieu, FR-75252 Paris Cedex 05, France
    \label{LPNHE}}
\titlefoot{Department of Physics, University of Lund,
     S\"olvegatan 14, SE-223 63 Lund, Sweden
    \label{LUND}}
\titlefoot{Universit\'e Claude Bernard de Lyon, IPNL, IN2P3-CNRS,
     FR-69622 Villeurbanne Cedex, France
    \label{LYON}}
\titlefoot{Dipartimento di Fisica, Universit\`a di Milano and INFN-MILANO,
     Via Celoria 16, IT-20133 Milan, Italy
    \label{MILANO}}
\titlefoot{Dipartimento di Fisica, Univ. di Milano-Bicocca and
     INFN-MILANO, Piazza della Scienza 3, IT-20126 Milan, Italy
    \label{MILANO2}}
\titlefoot{IPNP of MFF, Charles Univ., Areal MFF,
     V Holesovickach 2, CZ-180 00, Praha 8, Czech Republic
    \label{NC}}
\titlefoot{NIKHEF, Postbus 41882, NL-1009 DB
     Amsterdam, The Netherlands
    \label{NIKHEF}}
\titlefoot{National Technical University, Physics Department,
     Zografou Campus, GR-15773 Athens, Greece
    \label{NTU-ATHENS}}
\titlefoot{Physics Department, University of Oslo, Blindern,
     NO-0316 Oslo, Norway
    \label{OSLO}}
\titlefoot{Dpto. Fisica, Univ. Oviedo, Avda. Calvo Sotelo
     s/n, ES-33007 Oviedo, Spain
    \label{OVIEDO}}
\titlefoot{Department of Physics, University of Oxford,
     Keble Road, Oxford OX1 3RH, UK
    \label{OXFORD}}
\titlefoot{Dipartimento di Fisica, Universit\`a di Padova and
     INFN, Via Marzolo 8, IT-35131 Padua, Italy
    \label{PADOVA}}
\titlefoot{Rutherford Appleton Laboratory, Chilton, Didcot
     OX11 OQX, UK
    \label{RAL}}
\titlefoot{Dipartimento di Fisica, Universit\`a di Roma II and
     INFN, Tor Vergata, IT-00173 Rome, Italy
    \label{ROMA2}}
\titlefoot{Dipartimento di Fisica, Universit\`a di Roma III and
     INFN, Via della Vasca Navale 84, IT-00146 Rome, Italy
    \label{ROMA3}}
\titlefoot{DAPNIA/Service de Physique des Particules,
     CEA-Saclay, FR-91191 Gif-sur-Yvette Cedex, France
    \label{SACLAY}}
\titlefoot{Instituto de Fisica de Cantabria (CSIC-UC), Avda.
     los Castros s/n, ES-39006 Santander, Spain
    \label{SANTANDER}}
\titlefoot{Inst. for High Energy Physics, Serpukov
     P.O. Box 35, Protvino, (Moscow Region), Russian Federation
    \label{SERPUKHOV}}
\titlefoot{J. Stefan Institute, Jamova 39, SI-1000 Ljubljana, Slovenia
    \label{SLOVENIJA1}}
\titlefoot{Laboratory for Astroparticle Physics,
     University of Nova Gorica, Kostanjeviska 16a, SI-5000 Nova Gorica, Slovenia
    \label{SLOVENIJA2}}
\titlefoot{Department of Physics, University of Ljubljana,
     SI-1000 Ljubljana, Slovenia
    \label{SLOVENIJA3}}
\titlefoot{Fysikum, Stockholm University,
     Box 6730, SE-113 85 Stockholm, Sweden
    \label{STOCKHOLM}}
\titlefoot{Dipartimento di Fisica Sperimentale, Universit\`a di
     Torino and INFN, Via P. Giuria 1, IT-10125 Turin, Italy
    \label{TORINO}}
\titlefoot{INFN,Sezione di Torino and Dipartimento di Fisica Teorica,
     Universit\`a di Torino, Via Giuria 1,
     IT-10125 Turin, Italy
    \label{TORINOTH}}
\titlefoot{Dipartimento di Fisica, Universit\`a di Trieste and
     INFN, Via A. Valerio 2, IT-34127 Trieste, Italy
    \label{TRIESTE}}
\titlefoot{Istituto di Fisica, Universit\`a di Udine and INFN,
     IT-33100 Udine, Italy
    \label{UDINE}}
\titlefoot{Univ. Federal do Rio de Janeiro, C.P. 68528
     Cidade Univ., Ilha do Fund\~ao
     BR-21945-970 Rio de Janeiro, Brazil
    \label{UFRJ}}
\titlefoot{Department of Radiation Sciences, University of
     Uppsala, P.O. Box 535, SE-751 21 Uppsala, Sweden
    \label{UPPSALA}}
\titlefoot{IFIC, Valencia-CSIC, and D.F.A.M.N., U. de Valencia,
     Avda. Dr. Moliner 50, ES-46100 Burjassot (Valencia), Spain
    \label{VALENCIA}}
\titlefoot{Institut f\"ur Hochenergiephysik, \"Osterr. Akad.
     d. Wissensch., Nikolsdorfergasse 18, AT-1050 Vienna, Austria
    \label{VIENNA}}
\titlefoot{Inst. Nuclear Studies and University of Warsaw, Ul.
     Hoza 69, PL-00681 Warsaw, Poland
    \label{WARSZAWA}}
\titlefoot{Now at University of Warwick, Coventry CV4 7AL, UK
    \label{WARWICK}}
\titlefoot{Fachbereich Physik, University of Wuppertal, Postfach
     100 127, DE-42097 Wuppertal, Germany \\
\noindent
{$^\dagger$~deceased}
    \label{WUPPERTAL}}
\addtolength{\textheight}{-10mm}
\addtolength{\footskip}{5mm}
\clearpage

\headsep 30.0pt
\end{titlepage}

%
\pagenumbering{arabic}                              
\setcounter{footnote}{0}                            %
\large
\section{Introduction}
 
The Standard Model (SM), although in agreement with the available
experimental data~\cite{ewwg}, leaves several open questions. In
particular, the number of fermion generations and their mass spectrum
are not predicted.  The measurement of the ${\rm Z}$ decay
widths~\cite{ewwg} established that the number of light neutrino species
($m<m_{\rm Z}/2$, where $m_{\mathrm Z}$ is the ${\rm Z}$ boson mass) is
equal to three. However, if a heavy neutrino or a neutrinoless extra
generation exists, this bound does not exclude the possibility of extra
generations of heavy quarks. Moreover the fit to the electroweak
data~\cite{okun} does not deteriorate with the inclusion of one extra
heavy generation, if the new up and down-type quarks mass difference is
not too large. It should be noticed however that in this fit no mixing
of the extra families with the SM ones is assumed. 

The subject of this paper is the search for the pair production of a
fourth generation \blq at LEP-II: \bl production and decay are discussed
in section~\ref{intro}; in section~\ref{simula}, the data sets and the
Monte Carlo (MC) simulation are described; the analysis is discussed in
section~\ref{analysis}; the results and their interpretation within a
sequential model are presented in sections~\ref{results} and
\ref{const}, respectively.

\section{$\boldsymbol{\rm b'}$-quark production and decay \label{intro}}

Extra generations of fermions are predicted in several SM
extensions~\cite{frampton,djouadi}. In sequential
models~\cite{arhrib,hou,santos}, a fourth generation of fermions
carrying the same quantum numbers as the SM families is considered. In
the quark sector, an up-type quark, ${\rm t'}$, and a down-type quark,
${\rm b'}$, are included. The corresponding $4\times 4$ extended
Cabibbo-Kobayashi-Maskawa (CKM) matrix is unitary, approximately
symmetric and almost diagonal. As CP-violation is not considered in the
model, all the CKM elements are assumed to be real.

\begin{figure}
\vspace*{5mm}
\begin{center}
\begin{tabular}{c c c  c}
\includegraphics[width=6cm]{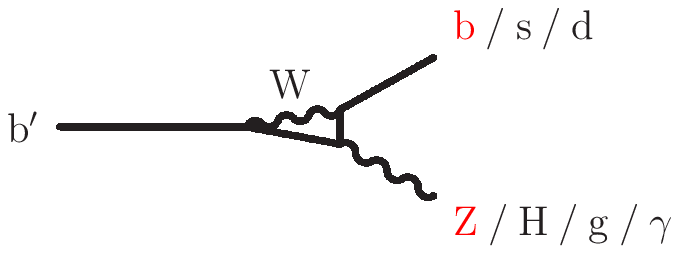} & & &
\includegraphics[width=6cm]{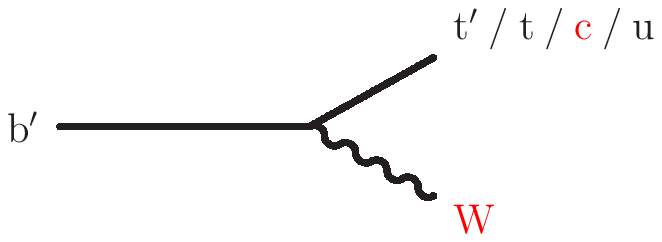} \\
& & &\\
a) & &  &b)\\
\end{tabular}
    \end{center} 

    \caption{{The Feynman diagrams corresponding  to the 
\bl (a) FCNC  and (b) CC decay modes are shown. } }
    \label{fig:decays}
\end{figure}

The \blq may decay via charged currents (CC) to \hbox{$U{\rm W}$}, with
$U={\rm t',t,c,u}$, or via flavour-changing neutral currents (FCNC) to $DX$,
where $D={\rm b,s,d}$ and $X={\rm Z,H,\gamma,g}$ (Fig.~\ref{fig:decays}). As in 
the SM, FCNC are absent at tree level, but can appear at one-loop level,
due to CKM mixing. If the \bl is lighter than ${\rm t'}$ and ${\rm t}$,
the decays ${\rm b'\to t'W}$ and ${\rm b'\to tW}$ are kinematically forbidden and
the one-loop FCNC decays can be as important as the CC
decays~\cite{hou}.

The analysis of the electroweak data~\cite{ewwg} shows that the mass
difference $|m_{\rm t'}-m_{\rm b'}|<60$~GeV$/c^2$ is consistent with the
measurement of the $\rho$ parameter~\cite{frampton,arhrib}. In particular,
when $m_{\rm Z}+m_{\rm b}<m_{\rm b'}<m_{\rm H}+m_{\rm b}$, either \bcw or
\bbz decay tend to be dominant~\cite{arhrib,hou,santos}.  In this case, the
partial widths of the CC and FCNC \bl decays depend mainly on $m_{\rm t'}$,
$m_{\rm b'}$ and $R_{CKM}=|\frac{V_{\rm cb'}} {V_{\rm tb'}V_{\rm tb}}|$,
where $V_{\rm cb'}$, $V_{\rm tb'}$ and $V_{\rm tb}$ are elements of the
extended $4\times 4$ CKM matrix~\cite{santos}.

Limits on the mass of the \blq have been set previously at various
accelerators. At LEP-I, all the experiments searched for \bl pair
production (${\rm e^+ e^-\to b'\bar{b'}}$), yielding a lower limit on
the \bl mass of about $m_{\rm Z}/2$~\cite{lep}. At the Tevatron, both
the D0\cite{d00} and CDF~\cite{cdf} experiments reported limits on
$\sigma({\rm p \bar p} \to {\rm b'} \bar{\rm b'})\times BR({\rm b' \to
bX})^2$, where $BR$ is the branching ratio corresponding to the
considered FCNC $b'$ decay mode and $X=\gamma,{\rm Z}$. Assuming
$BR({\rm b'}\to {\rm bZ})=1$, CDF excluded the region $100<m_{\rm
b'}<199$~GeV$/c^2$. Although no dedicated analysis was performed for the
\bcw decay, the D0 limits on $\sigma({\rm p \bar p} \to {\rm t} \bar{\rm
t})\times BR({\rm t \to cW})^2$ from Fig.~44 and Table~XXXI of
reference~\cite{d0} can give a hint on the possible values for
$BR({\mathrm b' \to cW})$\cite{pdg}.

In the present analysis the on-shell FCNC (${\rm b'\to bZ}$) and CC
(${\rm b'\to cW}$) decay modes were studied and consequently the mass
range $96$~GeV$/c^2 < m_{\rm b'} < 103$~GeV$/c^2$ was considered. This
mass range is complementary to the one covered by CDF~\cite{cdf}. The
mass range $m_{\rm W}+m_{\rm c} < m_{\rm b'} < m_{\rm Z}+m_{\rm b}$ was
not considered because in this region the evaluation of the branching
ratios for the different $\rm b'$ decays is particularly difficult from
the theoretical point of view~\cite{santos}. In the present analysis no
assumptions on the \brbz and $BR({\mathrm b' \to cW})$ in order to
derive mass limits were made.  Different final states, corresponding to
the different \bl decay modes and subsequent decays of the ${\rm Z}$ and
${\rm W}$ bosons, were analysed.

\section{Data samples and Monte Carlo simulation \label{simula}}

The analysed data were collected with the DELPHI detector~\cite{delphi}
during the years 1999 and 2000 in \hbox{LEP-II} runs at
$\sqrt{s}=196-209$~GeV and correspond to an integrated luminosity of
about 420~pb${}^{-1}$. The luminosity collected at each centre-of-mass energy
is shown in Table~\ref{lum}. During the year 2000, an unrecoverable
failure affected one sector of the central tracking detector (TPC),
corresponding to 1/12 of its acceptance. The data collected during the
year 2000 with the TPC fully operational were split into two energy
bins, below and above $\sqrt{s}=206$~GeV, with
$\langle\sqrt{s}\rangle=204.8$~GeV and
\hbox{$\langle\sqrt{s}\rangle=206.6$~GeV}, respectively. The data
collected with one sector of the TPC turned off were analysed separately
and have $\langle\sqrt{s}\rangle=206.3$~GeV.

\begin{table}[h!ptbc]
\begin{center}
\begin{tabular}{|c|c|c|c|c|c|c|}
\hline
$\sqrt{s}$ (GeV)& 196 & 200 & 202 & 205 & 207 & 206$^*$ 
\\
\hline
luminosity (pb$^{-1}$)& 76.0 &  82.7 & 40.2 & 80.0 & 81.9 & 59.2 \\   
\hline
\end{tabular}
\caption{The luminosity collected with the DELPHI detector at each 
centre-of-mass energy is shown. The energy bin labelled 206$^*$ 
corresponds to the data collected with one sector  of the TPC turned 
off.
}
\label{lum}
\end{center}
\end{table}

Signal samples were generated using a modified version of
PYTHIA~6.200~\cite{pythia}. Although PYTHIA does not provide FCNC decay
channels for quarks, it was possible to activate them by modifying the
decay products of an available channel. The angular distributions
assumed for \bl pair production and decay were those predicted by the SM
for any heavy down-type quark. Different samples, corresponding to \bl
masses in the range between 96 and 103~GeV$/c^2$ and with a spacing of
1~GeV$/c^2$ were generated at each centre-of-mass energy. Specific Monte Carlo 
simulations (for both SM and signal processes) were produced for the period when 
one sector of the TPC was turned off.

The most relevant background processes for the present analyses are those
leading to $WW$ or $ZZ$ bosons in the final state, \emph{i.e.}  four-fermion
backgrounds. Radiation in these events can mimic the six-fermion final
states for the signal. Additionally $q\bar q (\gamma)$ and Bhabha events can
not be neglected since for signal final states with missing energy these
backgrounds can become important. SM background processes were simulated at
each centre-of-mass energy using several Monte Carlo generators. All the
four-fermion final states (both neutral and charged currents) were generated
with WPHACT~\cite{wphact}, while the particular phase space regions of ${\rm
e^+ e^- \to e^+ e^- f \bar f}$ referred to as $\gamma\gamma$ interactions
were generated using PYTHIA~\cite{pythia}. The ${\rm qq(\gamma)}$ final state
was generated with KK2F~\cite{kk2f}. Bhabha events were generated with
BHWIDE~\cite{bhwide}.

The generated signal and background events were passed through the
detailed simulation of the DELPHI detector~\cite{delphi} and then
processed with the same reconstruction and analysis programs as the data. 

\section{Description of the analyses \label{analysis}}

\begin{figure}
\vspace*{5mm}
\begin{center}
\begin{tabular}{c c c c}
\includegraphics[width=6.3cm]{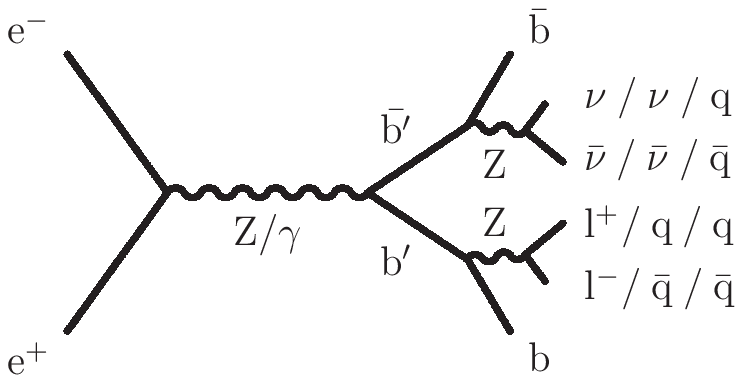} & & &
\includegraphics[width=5.8cm]{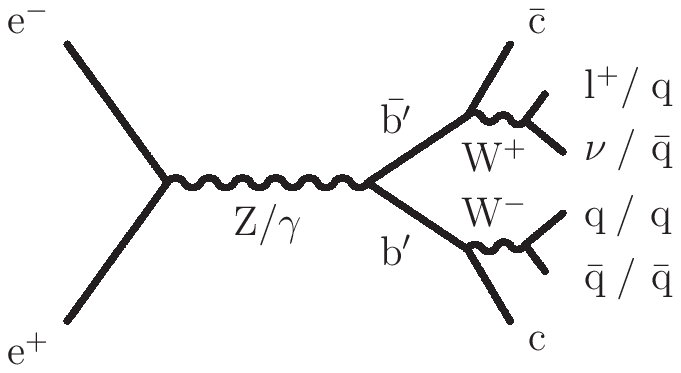} \\
& & &\\
a) & & & b)\\
\end{tabular}
    \end{center} 

    \caption{{The final states associated to the 
\bl (a) FCNC  and (b) CC decay modes are shown. Only those states analysed here are indicated.} }
    \label{fig:ncdecay}
\end{figure}

Pair production of \blq{}s was searched for in both the FCNC (${\rm b'\to
bZ}$) and CC (${\rm b'\to cW}$) decay modes. The \bl decay modes and the
subsequent decays of the gauge bosons (${\rm Z}$ or ${\rm W}$) lead to
several different final states (Fig.~\ref{fig:ncdecay}).  The
final states considered and their branching ratios are shown in
Table~\ref{topo}. The choice of the considered final states was done taking into 
account their signatures and BR. About 81\% and 90\% of the branching ratio to the 
FCNC and CC channels were covered, respectively.
All final states include two jets originating from the
low energy ${\rm b}$ (${\rm c}$) quarks present in the FCNC (CC) ${\rm
b'}$ decay modes. A common preselection was adopted, followed by a 
specific analysis for each of the final states (Table~\ref{topo}). 
\begin{center}
\begin{table}[h!]
\begin{center}
\begin{tabular}{|c || l | r | c |}
\hline
\bl decay & boson decays & $BR$ (\%) & final states \\
\hline
\hline
            &                                     &      &         \\
\bbz (FCNC) & ${\mathrm ZZ\to l^+ l^-\nu\bar\nu}$\hspace*{1mm} & 4.0\hspace*{4mm}  
& \bbllnn \\
            & ${\rm ZZ\to q\bar q\nu\bar\nu}$     & 28.0\hspace*{4mm} & \bbqqnn \\
            & ${\rm ZZ\to q\bar q q \bar q}$      & 48.6\hspace*{4mm} & \bbqqqq \\
\hline
\hline
            &                                     &      &         \\
\bcw (CC)   & ${\rm WW\to  q \bar ql^+ \nu}$\hspace*{1mm}  & 43.7\hspace*{4mm} 
& \ccqqln 
\\
            & ${\rm WW\to q \bar q q \bar q}$     & 45.8\hspace*{4mm} & \ccqqqq \\
\hline
\end{tabular}
\end{center}
\caption{{The final states considered in this analysis are shown.
About 81\% and 90\% of the branching ratio to the FCNC and CC channels 
were covered, respectively.
}}  
\label{topo}
\end{table}
\end{center}

Events were preselected by requiring at least eight good charged-particle
tracks and the visible energy measured at polar angles\footnote{In the
standard DELPHI coordinate system, the positive $z$ axis is along the
electron beam direction. The polar angle ($\theta$) is defined with respect
to the $z$ axis. In this paper, polar angle ranges are always assumed to be
symmetric with respect to the $\theta=90^\circ$ plane.} above $20^\circ$, to
be greater than $0.2\sqrt{s}$. Good charged-particle tracks were defined as
those with a momentum above 0.2~GeV$/c$ and impact parameters in the
transverse plane and along the beam direction below 4~cm and below 4~cm$/\sin
\theta$, respectively.

The identification of muons relied on the association of charged
particles to signals in the muon chambers and in the hadronic
calorimeters and was provided by standard DELPHI
algorithms~\cite{delphi}. The identification of electrons and photons
was performed by combining information from the electromagnetic
calorimeters and the tracking system. Radiation and interaction effects
were taken into account by an angular clustering procedure around the
main shower~\cite{remclu}.

The search for isolated particles (charged leptons and photons) was done 
by constructing double cones oriented in the direction of charged-particle 
tracks or neutral energy deposits. The latter ones were defined as 
calorimetric energy deposits above $0.5$~GeV, not matched to 
charged-particle tracks and identified as photon candidates by the 
standard DELPHI algorithms~\cite{delphi,remclu}. For charged leptons 
(photons), the energy in the region between the two cones, which had 
half-opening angles of $5^\circ$ and $25^\circ$ ($5^\circ$ and 
$15^\circ$), was required to be below 3~GeV (1~GeV), to ensure isolation. 
All the charged-particle tracks and neutral energy deposits inside the 
inner cone were associated to the isolated particle. Its energy was then 
re-evaluated as the sum of the energies inside the inner cone and was 
required to be above 5~GeV. For well identified leptons or 
photons~\cite{delphi,remclu} the above requirements were weakened. In this 
case only the external cone was used (to ensure isolation) and its angle 
$\alpha$ was varied according to the energy of the lepton (photon) candidate, 
down to $2^\circ$ for $P_\ell\geq 70$~GeV/$c$ ($3^\circ$ for $P_\gamma\geq 
90$~GeV/$c$), with the allowed energy inside the cone reduced by 
$\sin\alpha/\sin25^\circ$ ($\sin\alpha/\sin15^\circ$). Isolated leptons 
were required to have a momentum greater than 10 GeV$/c$ and a polar angle 
above $25^\circ$. Events with isolated photons were rejected.

All the events were clustered into two, four or six jets  using the Durham jet
algorithm~\cite{durham}, according to the number of jets expected in the 
signal in each of the final states, unless explicitly  stated otherwise.  
Although two ${\rm b}$ jets are 
always present in the FCNC final states, they have a relatively low energy and 
b-tagging techniques~\cite{btag} were not used.

Events were assigned to the different final states according to the
number of isolated leptons and to the missing energy in the event, as
detailed in Table~\ref{assig}. Within the same \bl decay channel, the
different selections were designed to be mutually exclusive. For the
final states involving charged leptons (\bbllnn and ${\mathrm c\bar c
q\bar q l^+\nu}$), events were divided into different samples
according to the lepton flavour identification: ${\rm e}$ sample (well
identified electrons), ${\rm \mu}$ sample (well identified muons) and
\emph{no-id} sample (leptons with unidentified flavour or two leptons
identified with different flavours).

Specific analyses were then performed for each of the final states. The
selection criteria for the \bbqqqq and \ccqqqq final states were the same.
The \bbllnn final state has a very clean signature (two leptons with $m_{\rm
l^+l^-}\sim m_{\rm Z}$, two low energy jets and missing mass close to $m_{\rm
Z}$) and consequently a sequential cut analysis was adopted. For all the
other final states, a sequential selection step was followed by a
discriminant analysis.  In this case, a signal likelihood (${\mathcal L}_S$)
and a background likelihood (${\mathcal L}_B$) were assigned to each event,
based on Probability Density Functions (PDF), built from the distributions of
relevant physical variables. The discriminant variable was defined as
$\ln({\mathcal L}_S/{\mathcal L}_B)$. 

\begin{center}
\begin{table}[t]
\begin{center}
\begin{tabular}{|c || c |}
\hline
final state & assignment criteria \\
\hline \hline
\bbllnn          & at least 1 isolated lepton \\
\hline
\bbqqnn          & no isolated leptons \\
   & $E_{missing}>50$~GeV \\
\hline
\bbqqqq    & no isolated leptons \\
           & $E_{missing}<50$~GeV \\
\hline \hline
\ccqqln    & only 1 isolated lepton \\
           & \\
\hline
\ccqqqq          & no isolated leptons \\
   & $E_{missing}<50$~GeV \\
\hline
\end{tabular}
 \end{center} 
\caption{{Summary of the final state assignment criteria.
}}
\label{assig}
\end{table}
\end{center}

\subsection{The \bdbbllnn final state}
 
\begin{figure}
 \begin{center}
  \Large DELPHI\\
  \vspace*{8mm}
  \includegraphics[width=14cm]{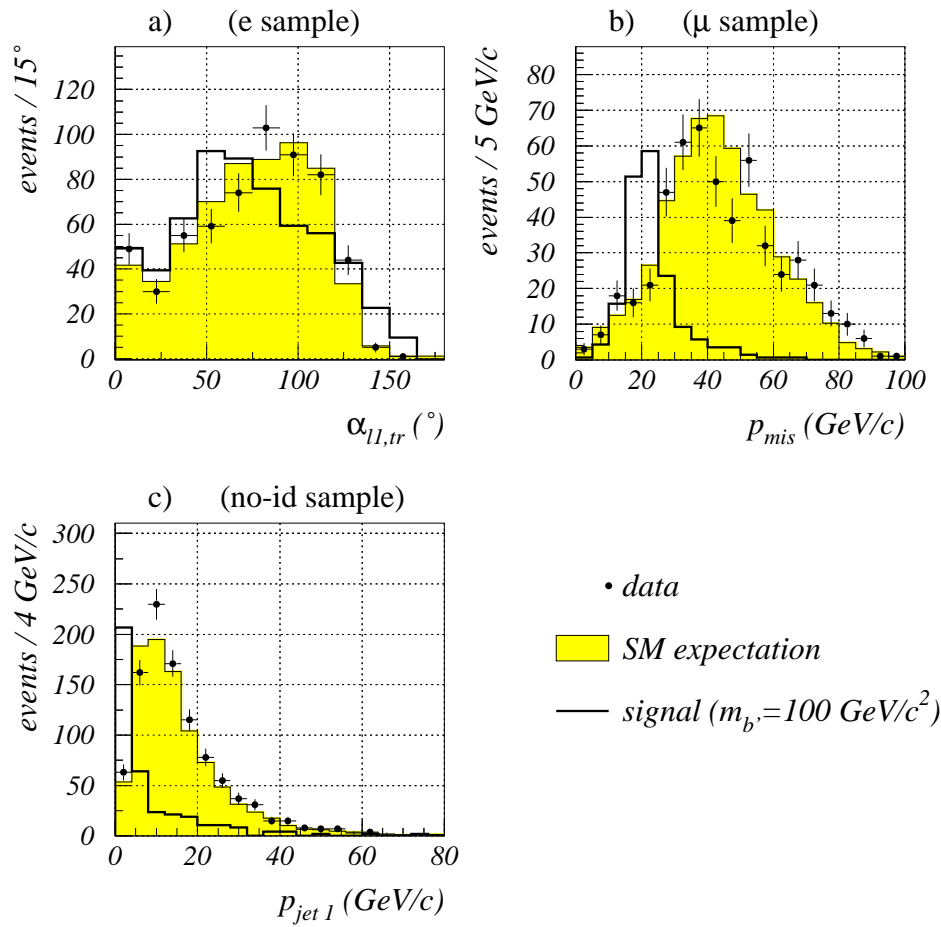}
 \end{center} 
 \caption{
Data and SM expectation  after the preselection
level for the \bbllnn final state and 
centre-of-mass energies above 200~GeV. 
(a) The angle between  the most energetic lepton and the closest 
charged-particle track (${\rm e}$ sample),
(b) the missing momentum ($\rm\mu$ sample) and 
(c) the momentum of the most energetic jet (\emph{no-id} sample) are shown.  
The signal distributions for \mblcem and $\sqrt{s}=205$~GeV are 
also shown with 
arbitrary normalisation. The background composition is 11\% of ${\rm q\bar 
q}$, 69\% of
${\rm WW}$, 15\% of ${\rm ZZ}$ and 5\% of $\gamma\gamma$ for the ${\rm e}$ 
sample, 6\% of ${\rm q\bar q}$,
90\% of ${\rm
WW}$ and 4\% of ${\rm ZZ}$ for the $\rm \mu$ sample and 45\% of ${\rm q\bar 
q}$,
48\% of ${\rm WW}$, 5\% of ${\rm ZZ}$ and 2\% of $\gamma\gamma$ for the 
\emph{no-id} sample.}
 \label{220sel0}
\end{figure}

\begin{figure}
 \begin{center}
  \Large DELPHI\\
  \vspace*{8mm}
  \includegraphics[width=14cm]{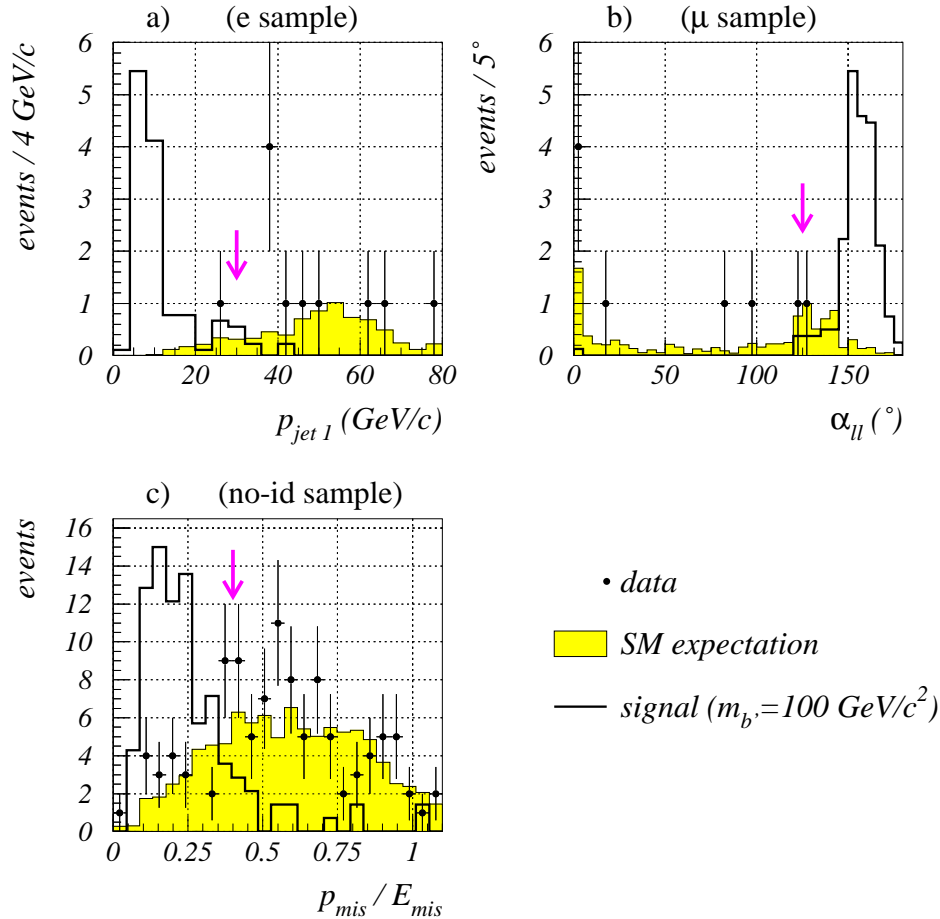}
 \end{center} 
 \caption{
Data and SM expectation after the first selection level
for the \bbllnn final state and for
centre-of-mass energies above 200~GeV. 
(a) The momentum of the most energetic jet (${\rm e}$ sample),
(b) the angle between the two leptons ($\rm\mu$ sample) and 
(c) the ratio between the missing momentum and missing energy (\emph{no-id}
sample) are shown.  The signal distributions for \mblcem and 
$\sqrt{s}=205$~GeV are also shown with arbitrary 
normalisation. 
The arrows represent the cuts applied in the second selection level.}
 \label{220sel2}
\end{figure}

The FCNC \bbllnn final state events were preselected as described above, by
requiring at least eight good charged-particle tracks, the visible energy
measured at polar angles above $20^\circ$, to be greater than $0.2\sqrt{s}$ and
at least one isolated lepton. Distributions of the relevant variables are shown
in Fig.~\ref{220sel0} for all the events assigned to this final state after the
preselection. The event selection was performed in two levels. In the first one,
events were required to have at least two leptons and an effective
centre-of-mass energy~\cite{sprime}, $\sqrt{s'}$, below $0.95\sqrt{s}$. The
particles other than the two leptons in the events were clustered into two jets
and the Durham resolution variable in the transition from two jets to one
jet\footnote{The Durham resolution variable is the minimum value of the scaled
transverse momentum obtained in the transition from $n$ to $n-1$
jets~\cite{durham} and will be represented by $y_{n \to n-1}$.} was required to
be greater than 0.002.  The number of data events and the SM expectation after
the first selection level is shown in Table~\ref{220}. The background
composition and the signal efficiencies at this level of selection for $m_{\rm
b'}=100$~GeV$/c^2$ and $\sqrt{s}=205$~GeV are given in Table~\ref{summary}. The
efficiencies for the other relevant \bl masses and $\sqrt{s}$ values were found
to be the same within errors. Data, SM expectation and signal distributions at
this selection level are shown in Fig.~\ref{220sel2}.

\begin{table}[h!]
\begin{center}
\begin{tabular}{|c || c  | c  | c |}
\hline
$\sqrt{s}$ (GeV)& \multicolumn{3}{c|}{ data (SM expectation $\pm$ statistical error)} \\
\hline
 & ${\rm e}$ sample & $\rm \mu$ sample & \emph{no-id} sample\\
\hline
\hline
196 & 2 (2.6$\pm$0.3) & 1 (2.9$\pm$0.3) & 47 (35.9$\pm$1.4) \\
\hline 
200 & 3 (2.5$\pm$0.4) & 4 (3.4$\pm$0.4) & 30 (37.4$\pm$1.4) \\
\hline 
202 & 2 (1.3$\pm$0.2) & 1 (1.7$\pm$0.2) & 20 (18.7$\pm$0.7)  \\
\hline 
205 & 5 (2.5$\pm$0.4) & 3 (3.0$\pm$0.4) & 35 (36.2$\pm$1.4) \\
\hline 
207 & 3 (2.3$\pm$0.4) & 3 (3.1$\pm$0.4) & 45 (35.1$\pm$1.3) \\
\hline 
206$^*$ & 1 (1.9$\pm$0.3) & 2 (2.6$\pm$0.2) & 31 (27.6$\pm$1.0) \\
\hline \hline
total & 16 (13.2$\pm$0.8) & 14 (16.7$\pm$0.8) & 208 (191.0$\pm$3.0) \\
\hline 
\end{tabular}
\caption{First selection level of the \bbllnn final state: the number 
of 
events selected in data and the
SM expectations after the first selection level for each sample and 
centre-of-mass energy are shown.}
\label{220}
\end{center}
\end{table}

In the final selection level the momentum of the more energetic (less energetic)
jet was required to be below 30~GeV$/c$ (12.5~GeV$/c$). Events in the ${\rm e}$
and \emph{no-id} samples had to have a missing energy greater than
$0.4\sqrt{s}$. In the $\rm \mu$ sample events were required to have an angle
between the two muons greater than 125$^\circ$. In the \emph{no-id} sample, the
angle between the two charged leptons had to be greater than 140$^\circ$ and
$p_{mis}/E_{mis} < 0.4$, where $p_{mis}$ and $E_{mis}$ are the missing momentum
and energy, respectively.  After the final selection, one data event was
selected for an expected background of 1.5$\pm$0.7.  This event belonged to the
\emph{no-id} sample and was collected at $\sqrt{s}=200$~GeV.  The signal
efficiencies for $m_{\rm b'}=100$~GeV$/c^2$ and $\sqrt{s}=205$~GeV are $30.6\pm
2.5$\% (${\rm e}$ sample), $48.6\pm 2.7$\% ($\rm \mu$ sample) and $7.2\pm 0.8$\%
(\emph{no-id} sample) and their variation with $m_{\rm b'}$ and $\sqrt{s}$ was
found to be negligible in the relevant range.

\subsection{ The \bdbbqqnn final state  \label{sec400}}

The FCNC \bbqqnn final state is characterised by the presence of four
jets and a missing mass close to $m_{\rm Z}$. At least 20 good
charged-particle tracks and $\sqrt{s'}>0.5\sqrt{s}$ were required.
Events were clustered into four jets. Monojet-like events were rejected
by requiring $-\log_{10}(y_{2\to 1})<0.7$ ($y_{2\to 1}$ is the Durham
resolution variable in the two to one jet transition). Furthermore,
$-\log_{10}(y_{4\to 3})$ was required to be below 2.8 and the energy of
the leading charged particle of the most energetic jet was required to
be below $0.1 \, \sqrt{s}$.

\begin{figure}
  \hspace*{3.5cm} \Large DELPHI\\
 \begin{center}
  \includegraphics[width=10cm]{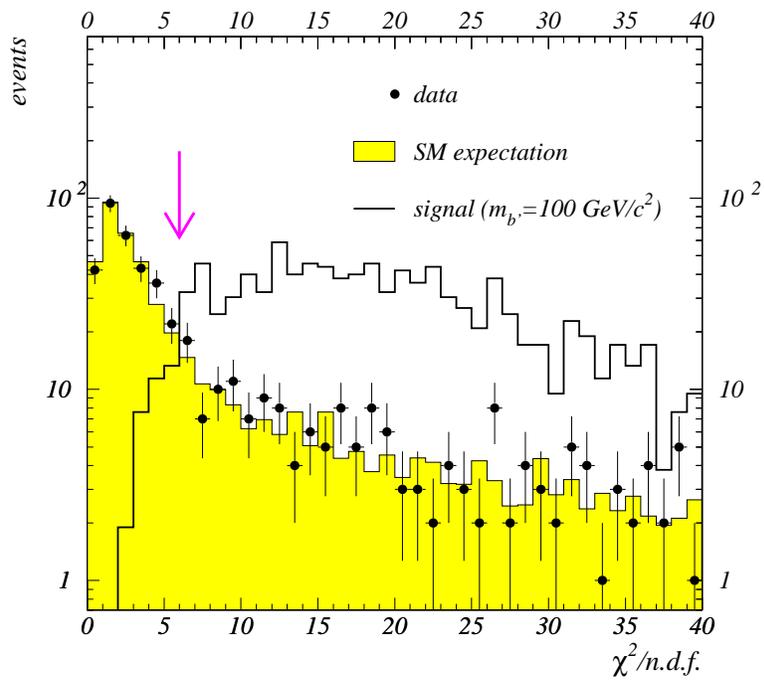}
 \end{center} 
 \caption{{Comparison of
data and SM expectation distributions of the $\chi^2/n.d.f.$ of the fit
imposing energy-momentum conservation and no missing energy 
for the \bbqqnn final state at centre-of-mass energies above 200~GeV. 
The arrow shows the applied cut. The signal for \mblcem
and $\sqrt{s}=205$~GeV is also shown with arbitrary normalisation.
}}
\label{p400}
\end{figure}

A kinematic fit imposing energy-momentum conservation and no missing energy was
applied and the background-like events with $\chi^2/n.d.f.<6$ were rejected. The
data, SM expectation and signal distributions of this variable are shown in
Fig.~\ref{p400}. Table~\ref{sel400} summarizes the number of selected data
events and the SM expectation. The background composition and the signal
efficiency at this level of selection for $m_{\rm b'}=100$~GeV$/c^2$ and
$\sqrt{s}=205$~GeV are given in Table~\ref{summary}. The efficiencies for the
other relevant \bl masses and $\sqrt{s}$ values were found to be the same within
errors.

\begin{table}[h!]
\begin{center}
\begin{tabular}{|c || c |}
\hline
$\sqrt{s}$ (GeV) & { data (SM expectation $\pm$ statistical error) } \\
\hline\hline
196  & 123 (106.3$\pm$4.0) \\
\hline
200  &111 (104.8$\pm$4.0) \\
\hline 
202  &50 (49.8$\pm$1.9) \\
\hline
205 &88 (94.2$\pm$3.7) \\
\hline
207  &99 (91.2$\pm$3.6) \\
\hline
206$^*$  &62 (65.7$\pm$2.6)\\
\hline
\hline
total & 533 (511.7$\pm$8.3) \\
\hline
\end{tabular}
\caption{First selection level of the \bbqqnn final state: the number of 
events 
selected in
data and the SM expectation
for each centre-of-mass energy are shown. 
}
\label{sel400}
\end{center}
\end{table}

\begin{figure}
 \begin{center}
  \Large DELPHI\\
  \vspace*{8mm}
  \includegraphics[width=14cm]{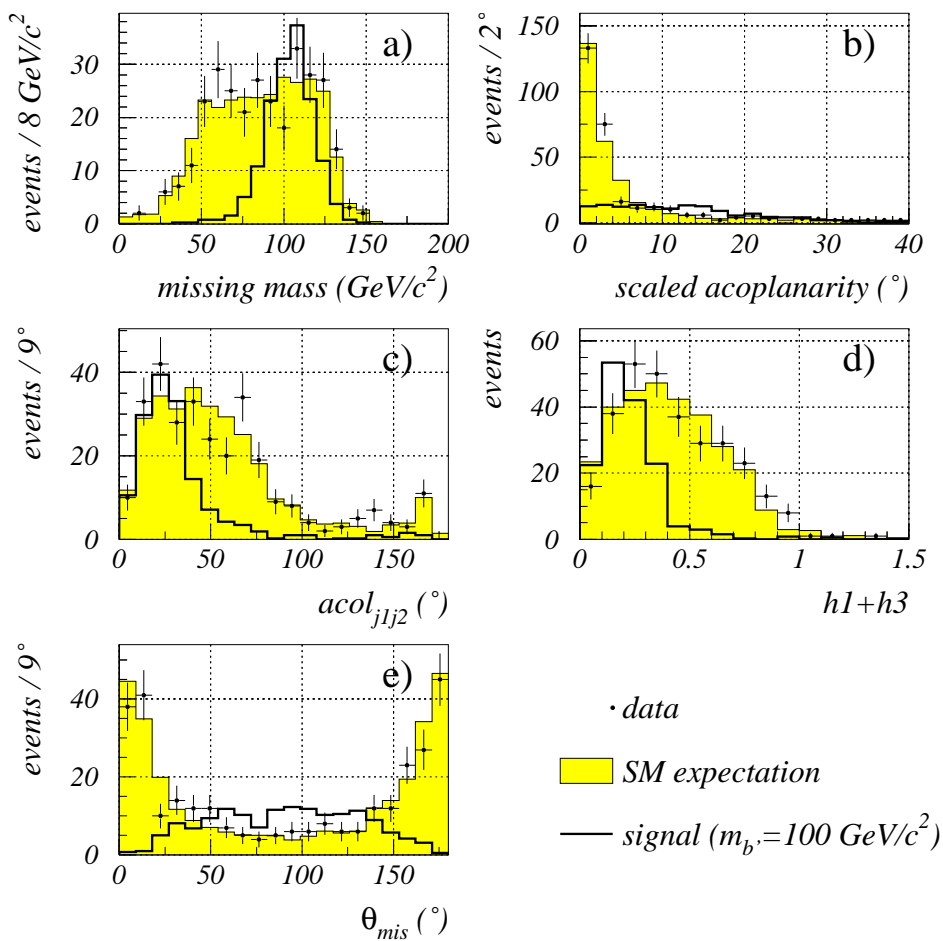}
 \end{center} 
 \caption{Variables used in the discriminant analysis (\bbqqnn final state). The 
data and SM expectation distributions for
centre-of-mass energies above 200~GeV are shown for
(a) the missing mass, (b)
$A^{j_1 j_2}_{cop} \times \mathrm{min}(\sin \theta_{j_1}, \sin
\theta_{j_2})$, where $A^{j_1 j_2}_{cop}$ is the acoplanarity
and $\theta_{j_1,j_2}$ are the polar angles of the jets when 
forcing the events into two jets,
(c) the acollinearity between the two
most energetic jets (with the event particles clustered into four jets), (d) the sum of the first and third 
Fox-Wolfram moments and
(e) the polar angle of the missing momentum.
The signal distributions for \mblcem and $\sqrt{s}=205$~GeV are also 
shown with arbitrary 
normalisation.
}
\label{f400}
\end{figure}

A discriminant selection was then performed using 
the following variables to build the PDFs:
\begin{itemize}
\item the missing mass;
\item $A^{j_1 j_2}_{cop} \times \mathrm{min}(\sin \theta_{j_1}, \sin
\theta_{j_2})$, where $A^{j_1 j_2}_{cop}$ is the acoplanarity\footnote{The
acoplanarity between two particles is defined as
$|180^\circ-|\phi_1-\phi_2||$, where $\phi_{1,2}$ are the azimuthal 
angles
of the two particles (in degrees).} and $\theta_{j_1,j_2}$ are the polar
angles of the jets when forcing the events into two jets\footnote{While
the signal is characterised by the presence of four jets in the final
state, the two jets configuration is used mainly for background 
rejection.};
\item the acollinearity between the two most energetic jets\footnote{The acollinearity
between two particles is defined as 
$180^\circ-\alpha_{1,2}$, where 
$\alpha_{1,2}$ is the angle (in degrees) between those two particles.} with the event particles
clustered into four jets;
\item the sum of the first and third Fox-Wolfram
moments ($h_1+h_3$)~\cite{fw};
\item the polar angle of the missing momentum.
\end{itemize}
The data, SM expectation and signal distributions of these variables are
shown in Fig.~\ref{f400}.

\subsection{The \bdbbqqqq final state \label{sec600}}

The FCNC \bbqqqq final state is characterised by the presence of six jets and a
small missing energy.  All the events were clustered into six jets and only
those with at least 30 good charged-particle tracks were accepted. Moreover,
events were required to have $\sqrt{s'}>0.6 \sqrt{s}$, $-\log_{10}(y_{2 \to 1})
< 0.7$ and $-\log_{10}(y_{6 \to 5}) < 3.6$. The number of selected data events
and the expected background at this level are shown in Table~\ref{600}. The
background composition and the signal efficiency at this level of selection for
$m_{\rm b'}=100$~GeV$/c^2$ and $\sqrt{s}=205$~GeV are given in
Table~\ref{summary}. The efficiencies for the other relevant \bl masses and
$\sqrt{s}$ values were found to be the same within errors.

\begin{table} [h!]
\begin{center}
\begin{tabular}{|c || c |}
\hline
$\sqrt{s}$ (GeV) & { data (SM expectation $\pm$ statistical error) } \\
\hline\hline
196 & 349 (326.7$\pm$5.3) \\
\hline
200 &347 (342.1$\pm$5.5) \\
\hline 
202 &165 (162.1$\pm$2.6) \\
\hline
205 &322 (319.0$\pm$5.2) \\
\hline
207 & 287 (307.6$\pm$5.0) \\
\hline
206$^*$ & 192 (215.8$\pm$3.6) \\
\hline
\hline
total & 1662 (1673.9$\pm$11.4) \\
\hline
\end{tabular}
\caption{First selection level of the \bbqqqq  and \ccqqqq final states: the number of 
events selected in
data and the SM expectations for each centre-of-mass energy are shown. 
}
\label{600}
\end{center}
\end{table}

\begin{figure}
 \begin{center}
  \Large DELPHI\\
  \vspace*{8mm}
  \includegraphics[width=14cm]{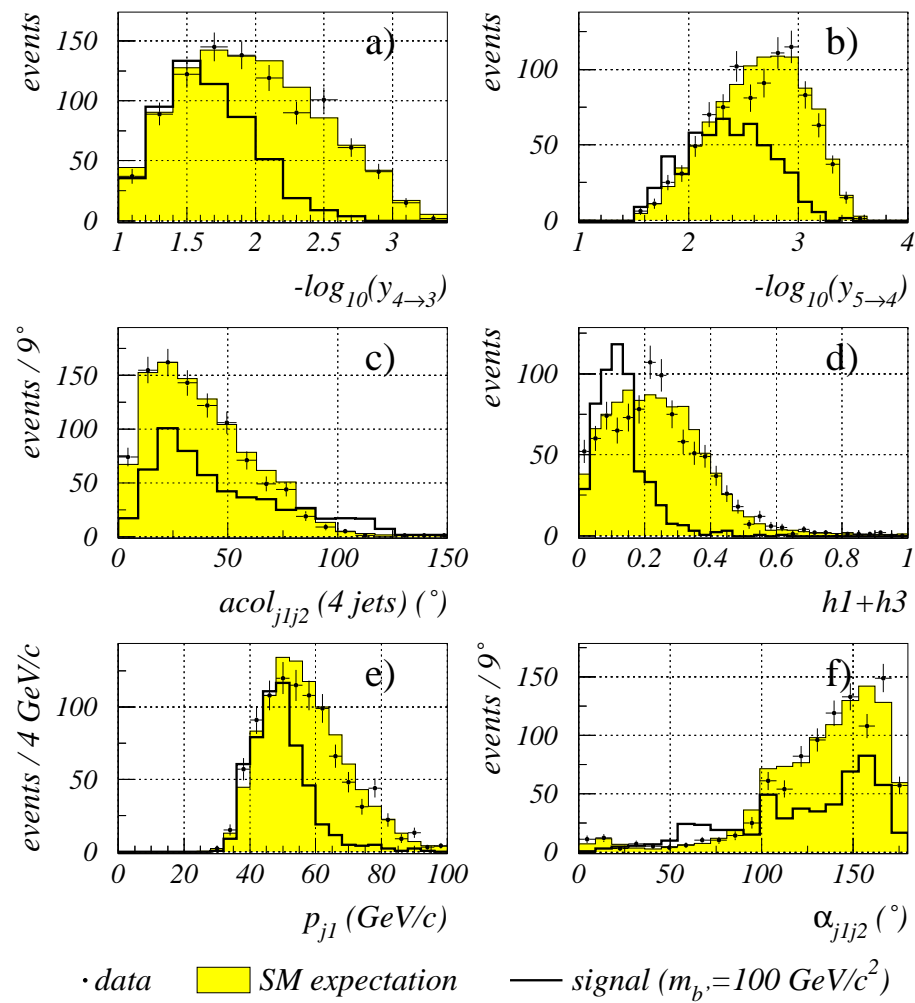}
 \end{center} 
 \caption{Variables used in the discriminant analysis (\bbqqqq final state). The 
data and SM  expectation for
centre-of-mass energies above 200~GeV are shown for (a) $-\log_{10} 
(y_{4\to 3})$, (b)
$-\log_{10} (y_{5\to 4})$, (c) the acollinearity between the two most
energetic jets, with the events clustered into four jets (see text for
explanation), (d) the $h1+h3$ Fox-Wolfram moments sum, (e) the momentum of
the most energetic jet and (f)  the angle between the two most energetic
jets. The signal distributions for \mblcem and
$\sqrt{s}=205$~GeV are also shown with arbitrary normalisation.
}
\label{f600}
\end{figure}

A discriminant selection was performed using the following variables to 
build the PDFs:
\begin{itemize}
\item the Durham resolution variable, $-\log_{10}(y_{4 \to 3})$;
\item the Durham resolution variable, $-\log_{10}(y_{5 \to 4})$;
\item the acollinearity between the two most energetic jets, with the 
event forced into four jets;
\item the sum of the first and third Fox-Wolfram moments;
\item the momentum of the most energetic jet;
\item the angle between the two most energetic jets (with the events 
clustered into six jets).
\end{itemize}
The distributions of these variables are
shown in Fig.~\ref{f600} for data, SM expectation and signal.

\subsection{The \bdccqqln final state}

The signature of this CC final state is the presence of four jets (two of them
having low energy), one isolated lepton and missing energy (originating from the
${\rm W\to l \bar\nu}$ decay). The events were accepted if they had at least 15
good charged-particle tracks. The event particles other than the identified
lepton were clustered into four jets. Part of the ${\rm q\bar q}$ and
$\gamma\gamma$ background was rejected by requiring $-\log_{10}(y_{2 \to 1}) <
0.7$.  Furthermore, there should be only one charged-particle track associated
to the isolated lepton, and the leading charged particle of the most energetic
jet was required to have a momentum below $0.1\sqrt{s}$.  The number of selected
data events and SM expectations at this level are summarized in
Table~\ref{4101}.  The background composition and the signal efficiencies at
this level of selection for $m_{\rm b'}=100$~GeV$/c^2$ and $\sqrt{s}=205$~GeV
are given in Table~\ref{summary}. The efficiencies for the other relevant \bl
masses and $\sqrt{s}$ values were found to be the same within errors.

\begin{table}[h!]
\vspace*{5mm}
\begin{center}
\begin{tabular}{|c || c | c | c|}
\hline
$\sqrt{s}$ (GeV) & \multicolumn{3}{c |}{data (SM expectation $\pm$ statistical error)} \\
\hline
 & ${\rm e}$ & $\rm\mu$ & \emph{no-id} \\ 
\hline\hline
196 & 65 (51.1$\pm$1.4) & 53 (56.1$\pm$1.5) & 38 (34.4$\pm$1.4) \\
\hline
200 & 54 (58.1$\pm$1.7) & 63 (59.9$\pm$1.6) & 40 (35.0$\pm$1.4)  \\
\hline 
202 & 30 (27.8$\pm$0.8) & 21 (28.4$\pm$0.8) & 13 (16.9$\pm$0.7)  \\
\hline
205 & 56 (50.8$\pm$1.5) & 66 (53.6$\pm$1.5) & 32 (33.3$\pm$1.4)  \\
\hline
207 & 53 (53.8$\pm$1.6) & 48 (57.2$\pm$1.6) & 35 (33.8$\pm$1.4)  \\
\hline
206$^*$ & 31 (37.2$\pm$1.4) & 42 (39.3$\pm$1.1) & 21 (23.4$\pm$1.0)  \\
\hline
\hline
total & 289 (278.8$\pm$3.5) & 293 (294.5$\pm$3.4) & 179 (176.8 $\pm$ 2.8) 
\\
\hline
\end{tabular}
\caption{First selection level of the \ccqqln final state: the number of 
events selected in
data and the SM expectations for each sample and
centre-of-mass energy are shown. 
}
\label{4101}
\end{center}
\end{table}

\begin{figure}
 \begin{center}
  \Large DELPHI\\
  \vspace*{8mm}
  \includegraphics[width=14cm]{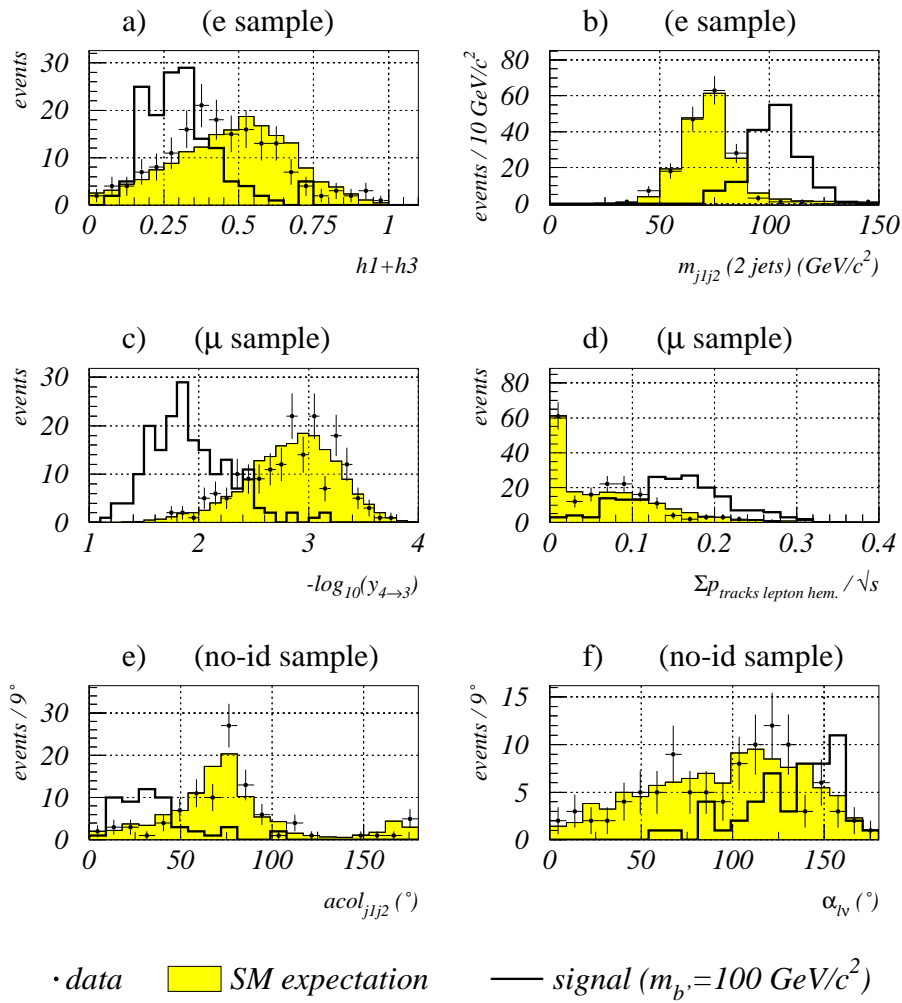}
 \end{center} 
 \caption{Variables used in the discriminant analysis (\ccqqln final state). 
The data events and background expectation for centre-of-mass energies 
above 200~GeV are shown for 
(a) the $h1+h3$ Fox-Wolfram moments sum ($\rm e$ sample), (b) the invariant mass 
of the two jets with the events clustered into two jets ($\rm e$ sample), (c)
$-\log_{10}(y_{4 \to 3})$ ($\rm\mu$ sample), (d) $\sum_i{|\vec 
p_i|}/\sqrt{s}$, 
where $\vec p_i$ are
the momenta of the charged particles (excluding the lepton) in the same 
hemisphere as the lepton 
($\rm\mu$ 
sample), (e) the acollinearity between the two most energetic jets 
(\emph{no-id} 
sample) and (f) the 
angle 
between the lepton
and the missing momentum (\emph{no-id} sample).  The signal 
distributions
for \mblcem and $\sqrt{s}=205$~GeV are also shown with arbitrary 
normalisation.
}
\label{f410x}
\end{figure}

The PDFs used to calculate the background and signal likelihoods 
were based on the following variables:
\begin{itemize}
\item the sum of the first and third Fox-Wolfram moments;
\item the invariant mass of the two jets, with the event particles 
other than the
identified lepton clustered into two jets;
\item the Durham resolution variable, $-\log_{10}(y_{4 \to 3})$;
\item  $\sum_i{|\vec p_i|}/\sqrt{s}$,
where $\vec p_i$ are 
the momenta of the charged particles (excluding the lepton) in the same 
hemisphere as the lepton (the hemisphere is defined with respect to the lepton);
\item  the acollinearity between the two most energetic jets;
\item the angle between the lepton and the missing momentum.
\end{itemize}
The data, SM expectation and signal distributions of these variables are
shown in Fig.~\ref{f410x}.

In order to improve the efficiency, events with no leptons seen in the detector
were kept in a fourth sample. For this sample, the selection criteria of the
\bbqqnn final state were applied and the same variables as in
section~\ref{sec400} were used to build the PDFs.  The signal efficiency after
the first selection level for \mblcem and $\sqrt{s}=205$~GeV was $8.9\pm
0.9$\%{}. The efficiencies for the other relevant \bl masses and $\sqrt{s}$
values were found to be the same within errors.

\subsection{The \bdccqqqq final state}

This final state is very similar to \bbqqqq (with slightly different kinematics
due to the mass difference between the ${\rm Z}$ and the ${\rm W}$). The
analysis described in section~\ref{sec600} was thus adopted. The number of
selected events and the SM expectations can be found in Table~\ref{600}.  At
this level, the signal efficiency for \mblcem and $\sqrt{s}=205$~GeV was
$67.3\pm 1.5$\%{}. The efficiencies for the other \bl masses and centre-of-mass
energies were the same within errors. The PDFs were built using the same set of
variables as in section~\ref{sec600}.

\section{Results \label{results}}

\begin{table}[h!ptbc]
\begin{center}
\begin{tabular}{|c r||c|c c c c| c |}
\hline
                           &        & data & \multicolumn{4}{|c|}{background} & signal \\ 
\multicolumn{2}{|c||}{final state} & (SM $\pm$ stat. error) & \multicolumn{4}{|c|}{composition (\%{}) } & efficiency  (\%{})\\
 & & & ${\rm q\bar q}$ & ${\rm WW}$ & ${\rm ZZ}$ & $\gamma\gamma$ & \\
\hline
\hline
\bbllnn            & e sample         & 16 (13.2$\pm$0.8)                    & 16 & 16 & 68 & 0 & 35.1$\pm$2.6\\
(first selection   & $\mu$ sample     & 14 (16.7$\pm$0.8)                    & 0  & 10 & 90 & 0 & 53.4$\pm$2.7\\
level)             & \phantom{aaa} \emph{no-id} sample & 208 (191.0$\pm$3.0) & 8  & 80 & 12 & 0 & 12.3$\pm$1.0\\
\hline
{\bbqqnn} & & 533 (511.7$\pm$8.3)                                            & 76 & 17 & 2  & 5 & 57.6$\pm$1.7\\
\hline
{\bbqqqq} & & 1662 (1673.9$\pm$11.4)                                         & 35 & 65 & 0  & 0 & 66.0$\pm$1.5\\
\hline
 & e sample         & 289 (278.8$\pm$3.5)                                    & 7  & 82 & 11 & 0 & 45.3$\pm$2.7\\
\ccqqln & $\mu$ sample  & 293 (294.5$\pm$3.4)                                & 2  & 97 & 1  & 0 & 56.4$\pm$2.7\\  
 & \phantom{a} \emph{no-id} sample & 179 (176.8$\pm$2.8)                     & 9  & 84 & 7  & 0 &  5.3$\pm$0.7\\
 & no lepton sample & 533 (511.7$\pm$8.3)                                    & 76 & 17 & 2  & 5 &  8.9$\pm$0.9\\
\hline 
{\ccqqqq } & & 1662 (1673.9$\pm$11.4)                                        & 35 & 65 & 0  & 0 & 67.3$\pm$1.5\\
\hline
\end{tabular}
\caption{Summary of the total number of selected data events and SM
expectations for the studied final states after the final selection (first
selection level for \bbllnn ). The corresponding background composition and
signal efficiencies for \mblcem and $\sqrt{s}=205$~GeV are also shown.}
\label{summary}
\end{center}
\end{table}

\begin{figure}
 \begin{center}
  \Large DELPHI\\
  \includegraphics[width=14cm]{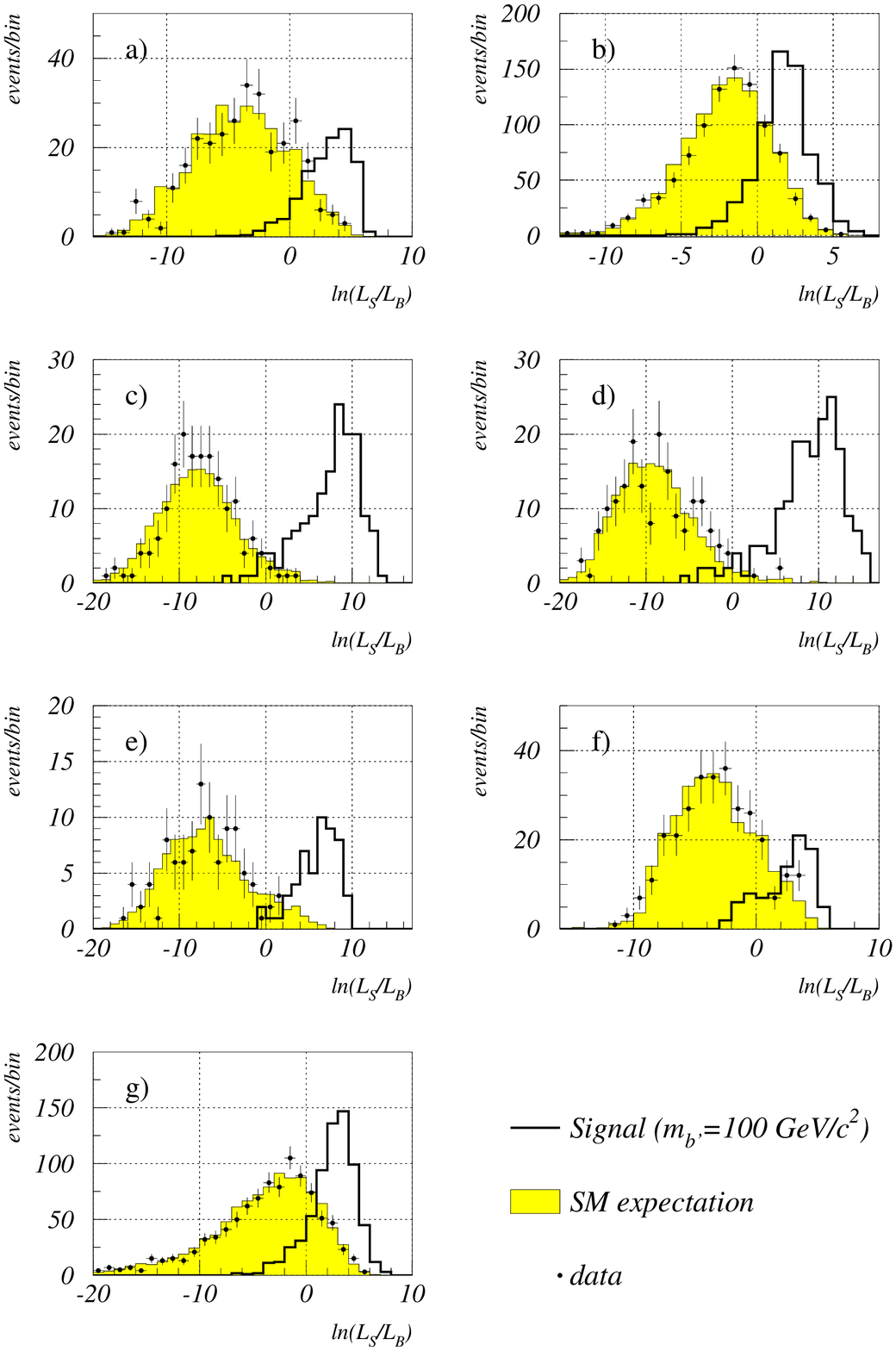}
 \end{center} 
 \caption{{Discriminant variables $\ln({\mathcal L}_S/{\mathcal L}_B)$
for data and SM simulation (centre-of-mass energies above 200~GeV). FCNC 
$\rm b'$ decay mode: (a) 
\bbqqnn and (b) ${\rm b\bar b q\bar qq\bar q}$. 
CC $b'$ decay mode: (c) \ccqqln (${\rm e}$ sample), (d) 
\ccqqln ($\rm\mu$ sample),
(e) \ccqqln (\emph{no-id} sample) (f) \ccqqln (\emph{no
lepton} sample) and (g) ${\rm c \bar cq \bar qq \bar q}$. 
The signal distributions for
\mblcem and $\sqrt{s}=205$~GeV are also shown with arbitrary 
normalisation.
}}
\label{like}
\end{figure}

For all final states, a good agreement between data and SM expectation was
found. The summary of the total number of selected data events, SM
expectations, the corresponding background composition and the signal
efficiencies for the studied final states are shown in Table~\ref{summary}.
In the \bbllnn final state, one data event was retained after the final
selection level, for a SM expectation of 1.5~$\pm$~0.7 events. This event
belonged to the \emph{no-id} sample and was collected at $\sqrt{s}=200$~GeV. 
For all the other final states, discriminant analyses were used. In these
cases, a discriminant variable, $\ln({\mathcal L}_S/{\mathcal L}_B)$, was
defined. The distributions of $\ln ({\mathcal L}_S/{\mathcal L}_B)$, for the
different analysis channels are shown in Fig.~\ref{like}.  No evidence for a
signal was found in any of the channels and the full information,
\emph{i.e.} event numbers and the shapes of the distributions of the
discriminant variables were used to derive limits on \brbz and $BR({\mathrm
b' \to cW})$.

\subsection{Limits on \bdbrbz and \bdbrcw}

Upper limits on the product of the ${\rm e^+ e^- \to b' \bar{\rm b'}}$
cross-section and the branching ratio as a function of the \bl mass were
derived at 95\% confidence level (CL) in each of the considered \bl decay
modes (FCNC and CC), taking into account the values of the discriminant
variables and their expected distributions for signal and background, the
signal efficiencies and the data luminosities at the various
centre-of-mass energies.

\begin{figure}
 \begin{center}
  \Large DELPHI\\
  \vspace*{-3mm}
  \includegraphics[width=15cm]{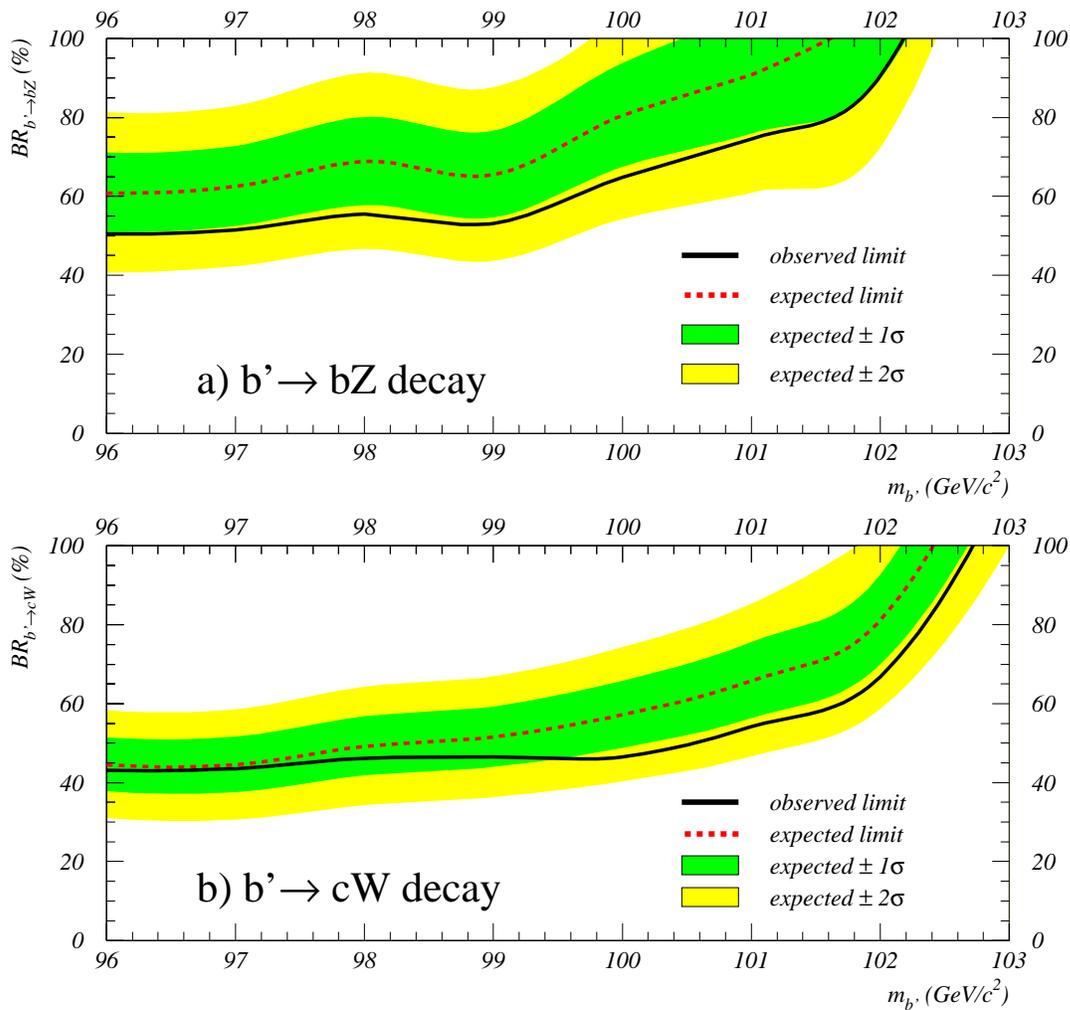}
 \end{center} 
 \caption{{The observed and 
expected 
upper limits at 95\% CL
 on  (a) \brbz 
and (b) \brcw are shown. The $1 \sigma$ and $2 \sigma$
bands around the expected limit are also presented. Systematic
errors were taken into account in the limit evaluation.
}}
\label{limbr}
\end{figure}

Assuming the SM cross-section for the pair production of heavy quarks at
LEP\cite{santos,pythia}, these limits were converted into limits on the
branching ratios corresponding to the \bbz and \bcw decay modes. The
modified frequentist likelihood ratio method~\cite{ar} was used.  The
different final states and centre-of-mass energy bins were treated as
independent channels. For each \bl mass only the channels with
$\sqrt{s}>2\,m_{\rm b'}$ were considered. In order to avoid some
non-physical fluctuations of the distributions of the discriminant
variables due to the limited statistics of the generated events, a
smoothing algorithm was used. The median expected limit, \emph{i.e.} the
limit obtained if the SM background was the only contribution in data,
was also computed. In Fig.~\ref{limbr} the observed and expected limits
on \brbz and \brcw are shown as a function of the \bl mass. The
$1\sigma$ and $2\sigma$ bands around the expected limit are also shown.
The observed and expected limits are statistically compatible.  At 95\%
CL and for $m_{\rm b'}=96$~GeV$/c^2$, the $BR({\mathrm b' \to bZ})$ and
$BR({\mathrm b' \to cW})$ have to be below $51$\% and $43$\%{},
respectively. These limits were evaluated taking into account the
systematic uncertainties, as explained in the next subsection.

The limits obtained for $BR({\mathrm b' \to bZ})$ are compatible with
those presented by CDF~\cite{cdf} for a ${\rm b'}$ mass of
100~GeV$/c^2$. Below this mass, the DELPHI result is more sensitive and
the CDF limit degrades rapidly. For higher ${\rm b'}$ masses, the LEP-II
kinematical limit is reached and the present analysis looses
sensitivity.

\subsection{Systematic uncertainties}

The evaluation of the limits was performed taking into account systematic
uncertainties, which affect the background estimation, the signal efficiency
and the shape of the distributions used. The following systematic 
uncertainties
were considered:
\begin{itemize} 
\item
SM cross-sections: uncertainties on the
SM cross-sections translate into uncertainties on the expected
number of background events. The overall uncertainty on the most
relevant SM background processes for the present analyses is typically
less than 2\%~\cite{lep2}, which leads to relative changes on the
branching ratio limits below 6\%{};

\item Signal generation: uncertainties on the final state quark
hadronisation and fragmentation modelling were studied. The Lund
symmetric fragmentation function was tested and compared with schemes
where the ${\rm b}$ and ${\rm c}$ quark masses are taken into
account~\cite{pythia}. This systematic error source was estimated to be
of the order of 20\% in the signal efficiency, by conservatively taking
the maximum observed variation. The relative effect on the branching
ratio limits is below 16\%{};

\item Smoothing: the uncertainty associated to the
discriminant variables smoothing was estimated by applying different 
smoothing algorithms. The
smoothing procedure does not change the number of SM expected events or the
signal efficiency, but may lead to differences in the shape of the
discriminant variables. The relative effect of this uncertainty on the
limits evaluation was found to be below 9\%{}.  
\end{itemize} 
Further details on the evaluation of the systematic errors and the 
derivation of limits can be found in~\cite{tese}.

\section{Constraints on $\boldsymbol{R_{CKM}}$ \label{const}}

The branching ratios for the \bl decays can be computed within a four
generations sequential model~\cite{arhrib,hou,santos}. As discussed before,
if the \bl is lighter than both the ${\rm t}$ and the ${\rm t'}$ quarks and 
$m_{\rm Z} \, <
m_{\rm b'} < m_{\rm H}$, the main contributions to the \bl width are \brbz and 
\brcw\cite{santos}. 
Using the unitarity of the CKM matrix, its approximate diagonality ($V_{\rm ub'} 
\,
V_{\rm ub} \approx \, 0$) and taking $V_{\rm cb} \approx \, 10^{-2}$~\cite{pdg}, 
the
branching fractions can be written as a function
of three variables: $R_{CKM}=|\frac{V_{\rm cb'}}{V_{\rm tb'} \, V_{\rm tb}}|$, 
$m_{\rm t'}$
and $m_{\rm b'}$~\cite{arhrib,hou,santos}.

\begin{figure}
\hspace*{4cm}\Large DELPHI
\vspace*{-3mm}
 \begin{center}
  \includegraphics[width=10cm]{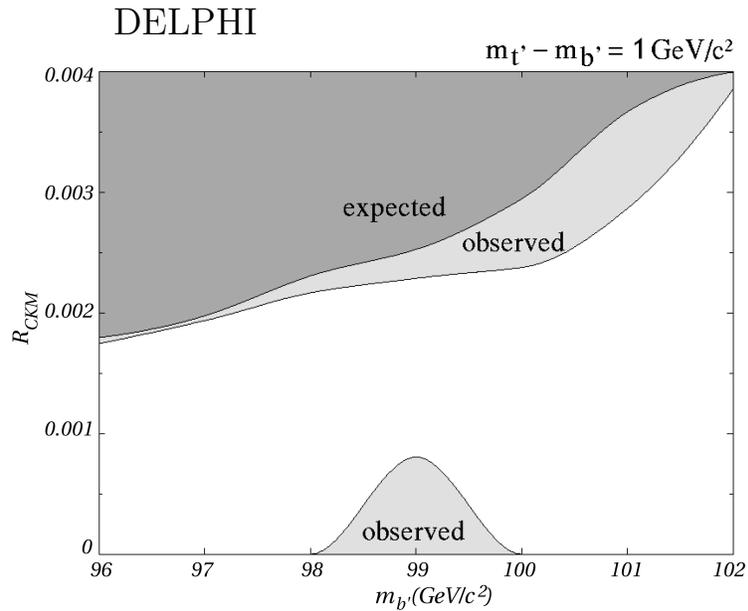}
 \end{center} 
 \caption{{The
excluded region in the plane ($R_{CKM}$, $m_{\rm b'}$) with
$m_{\rm t'}-m_{\rm b'}=1$~GeV$/c^2$, obtained
from the 95\% CL upper limits on \brbz (bottom)
and \brcw (top) is shown. The light and dark
shadings correspond to the observed and
expected limits, respectively. The expected limits on \brbz did not
allow exclusions to be set for low values of $R_{CKM}$.
}}
\label{lip1}
\end{figure}

\begin{figure}
\hspace*{4cm}\Large DELPHI
\vspace*{-3mm}
 \begin{center}
  \includegraphics[width=10cm]{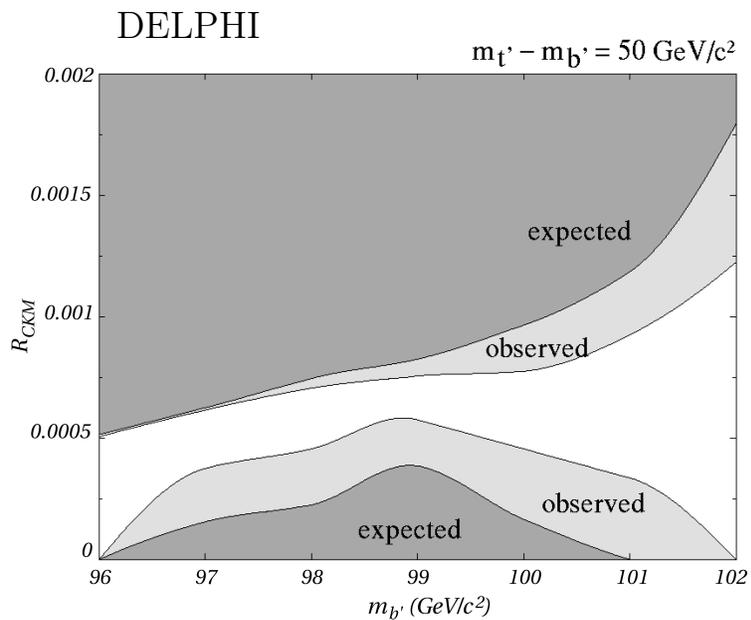}
 \end{center} 
 \caption{{The
excluded region in the plane ($R_{CKM}$, $m_{\rm b'}$) with
$m_{\rm t'}-m_{\rm b'}=50$~GeV$/c^2$, obtained
from the 95\% CL upper limits on \brbz (bottom)  
and \brcw (top) is shown. The light and dark
shadings correspond to the observed and
expected limits, respectively.
}}
\label{lip50}
\end{figure}

Fixing $m_{\rm t'}-m_{\rm b'}$, the limits on \brbz and \brcw
(Fig.~\ref{limbr}) can be translated into 95\% CL bounds on $R_{CKM}$ as a
function of $m_{\rm b'}$. Two extreme cases were considered: the almost
degenerate case, with $m_{\rm t'}-m_{\rm b'}=1$~GeV$/c^2$, and the case in which 
the
mass difference is close to the largest possible value,
$m_{\rm t'}-m_{\rm b'}=50$~GeV$/c^2$~\cite{frampton,arhrib}. The results are shown
in Fig.~\ref{lip1} and Fig.~\ref{lip50}. In the figures, the upper curve was
obtained from the limit on $BR({\mathrm b' \to cW})$, while the lower curve was
obtained from the limit on $BR({\mathrm b' \to bZ})$, which decreases with growing
$m_{\rm t'}$. 
This suppression is due to the GIM mechanism~\cite{gim} as $m_{\rm t'}$
approaches $m_{\rm t}$. On the other hand, as the \bl mass approaches 
the ${\rm bZ}$ threshold, the
${\rm b'\to bg}$ decay dominates over ${\rm b'\to bZ}$~\cite{santos} and the lower 
limit on $R_{CKM}$ becomes less stringent.
The expected limits on $BR({\mathrm b' \to bZ})$
did not allow to set exclusions for low values of $R_{CKM}$ and
$m_{\rm t'}-m_{\rm b'}=1$~GeV$/c^2$ (see Fig.~\ref{lip1}).

\section{Conclusions}

The data collected with the DELPHI detector at $\sqrt{s}=196-209$~GeV show
no evidence for the pair production of \blq{}s with masses ranging
from 96 to 103~GeV$/c^2$.

Assuming the SM cross-section for the pair production of heavy quarks at
LEP, 95\% CL upper limits on \brbz and \brcw were obtained. It was shown
that, at 95\% CL and for $m_{\rm b'}=96$~GeV$/c^2$, the $BR({\mathrm b'
\to bZ})$ and $BR({\mathrm b' \to cW})$ have to be below $51$\% and
$43$\%{}, respectively. The 95\% CL upper limits on the branching
ratios, combined with the predictions of the sequential fourth
generation model, were used to exclude regions of the ($R_{CKM}$,
$m_{\rm b'}$) plane for two hypotheses of the $m_{\rm t'}-m_{\rm b'}$
mass difference. It was shown that, for $m_{\rm t'}-m_{\rm
b'}=1~(50)$~GeV$/c^2$ and $96$~GeV$/c^2 < m_{\rm b'} < 102$~GeV$/c^2$,
$R_{CKM}$ is bounded by an upper limit of $3.8\times 10^{-3}$ 
($1.2\times
10^{-3}$). For $m_{\rm b'}=100$~GeV$/c^2$ and $m_{\rm t'}-m_{\rm
b'}=50$~GeV$/c^2$, the CKM ratio was constrained to be in the range
$4.6\times 10^{-4}<R_{CKM}<7.8\times 10^{-4}$.


\newpage
\subsection*{Acknowledgements}
\vskip 3 mm
We are greatly indebted to our technical 
collaborators, to the members of the CERN-SL Division for the excellent 
performance of the LEP collider, and to the funding agencies for their
support in building and operating the DELPHI detector.\\
We acknowledge in particular the support of \\
Austrian Federal Ministry of Education, Science and Culture,
GZ 616.364/2-III/2a/98, \\
FNRS--FWO, Flanders Institute to encourage scientific and technological 
research in the industry (IWT) and Belgian Federal Office for Scientific, 
Technical and Cultural affairs (OSTC), Belgium, \\
FINEP, CNPq, CAPES, FUJB and FAPERJ, Brazil, \\
Czech Ministry of Industry and Trade, GA CR 202/99/1362,\\
Commission of the European Communities (DG XII), \\
Direction des Sciences de la Mati$\grave{\mbox{\rm e}}$re, CEA, France, \\
Bundesministerium f$\ddot{\mbox{\rm u}}$r Bildung, Wissenschaft, Forschung 
und Technologie, Germany,\\
General Secretariat for Research and Technology, Greece, \\
National Science Foundation (NWO) and Foundation for Research on Matter (FOM),
The Netherlands, \\
Norwegian Research Council,  \\
State Committee for Scientific Research, Poland, SPUB-M/CERN/PO3/DZ296/2000,
SPUB-M/CERN/PO3/DZ297/2000, 2P03B 104 19 and 2P03B 69 23(2002-2004)\\
FCT - Funda\c{c}\~ao para a Ci\^encia e Tecnologia, Portugal, \\
Vedecka grantova agentura MS SR, Slovakia, Nr. 95/5195/134, \\
Ministry of Science and Technology of the Republic of Slovenia, \\
CICYT, Spain, AEN99-0950 and AEN99-0761,  \\
The Swedish Research Council,      \\
Particle Physics and Astronomy Research Council, UK, \\
Department of Energy, USA, DE-FG02-01ER41155, \\
EEC RTN contract HPRN-CT-00292-2002. \\


\end{document}